\documentclass[journal]{IEEEtran}

\usepackage[dvips]{graphicx}
\graphicspath{{./}}
\DeclareGraphicsExtensions{.eps}
\usepackage[cmex10]{amsmath}
\begin{document}

\title{A* Based Algorithm for Reduced Complexity ML Decoding of Tailbiting Codes}

\author{Jorge~Ort\' in,~\IEEEmembership{Student~Member,~IEEE,}
       Paloma~Garc\' ia,~Fernando~Guti\' errez~and~Antonio~Valdovinos%
\thanks{The authors are with the Arag\' on Institute for Engineering Research (I3A), University of Zaragoza, Zaragoza E-50018, Spain (e-mail: jortin@unizar.es).

This work has been financed by the Spanish Government (FPU grant to the first author and Project TEC2008-06684-C03-02/TEC from MCI and FEDER), Gobierno de Arag\'on (Project PI003/08 and WALQA Technology Park) and the European IST Project EUWB.}}

\maketitle

\begin{abstract}
The A* algorithm is a graph search algorithm which has shown good results in terms of computational complexity for Maximum Likelihood (ML) decoding of tailbiting convolutional codes. The decoding of tailbiting codes with this algorithm is performed in two phases. In the first phase, a typical Viterbi decoding is employed to collect information regarding the trellis. The A* algorithm is then applied in the second phase, using the information obtained in the first one to calculate the heuristic function. The improvements proposed in this work decrease the computational complexity of the A* algorithm using further
information from the first phase of the algorithm. This information is used for obtaining a more accurate heuristic function and finding early terminating conditions for the A* algorithm. Simulation results show that the proposed modifications decrease the complexity of ML decoding with the A* algorithm in terms of the performed number of operations.
\end{abstract}

\begin{IEEEkeywords}
Convolutional codes, decoding, tailbiting, A* algorithm, IA* algorithm.
\end{IEEEkeywords}

\section{Introduction}
\IEEEPARstart{T}{ailbiting} convolutional codes are formed when the encoder memory is preset with the last bits of the block being encoded. This fact ensures that the initial and final states of the trellis are the same and avoids the presence of the zero tail at the end of the block, hence improving the efficiency of the code, especially for short block sizes. For this reason, tailbiting codes are employed in new cellular radio systems such as Wimax or LTE. Nevertheless, the complexity of the decoding process is also increased, since the decoder does not know the value of the initial and final states of the trellis.

An evident algorithm to obtain a Maximum Likelihood (ML) decoding of each encoded block would be to perform a different Viterbi decoding for each of the possible initial and
final states of the trellis and select the decoded path with the best metric. However, this algorithm is impractical in most cases due to its large computational complexity.

As a consequence, alternative suboptimal decoding algorithms have been proposed, many of them based on the Circular Viterbi Algorithm (CVA) \cite{ref1}. These algorithms use
the circular property of tailbiting codes, employing the metrics accumulated in the ending states of the trellis as the initial ones for a new Viterbi decoding until a termination condition is fulfilled. Amongst these algorithms, the one proposed in \cite{ref2} obtains the best results in terms of both BER and computational load.

Recently, a new approach to ML decoding of tailbiting codes was proposed based on the A* algorithm \cite{ref3}. The A* algorithm is a search algorithm which finds the least cost path
from an initial node in a graph to a goal node optimizing a defined function. In this sense, the A* algorithm can achieve ML decoding in a tailbiting trellis since it always obtains the
path with the best metric from a selected initial state to a final state. Amongst the decoding methods based on the A* algorithm, the IA* algorithm proposed in \cite{ref4} and its updated version \cite{ref5} achieve the best known results in terms of the
required branch metric calculations to perform the decoding, as its computational complexity is lower than that of the suboptimal algorithms based on the CVA.

In this work, we introduce two new modifications on the use of the A* algorithm to decode tailbiting convolutional codes which further decrease the computational complexity of
the decoding process. The first improvement uses information regarding the survivor paths found in the first phase of the algorithm for finding early terminating conditions in the
application of the A* algorithm in the second phase.

Additionally, the second modification improves the estimation of the distance, measured in terms of accumulated metric, from the initial state of a subtrellis to its end, employing the
information of the discarded paths obtained in the application of the Viterbi algorithm in the first phase. This estimation is employed in the heuristic function of the A* algorithm. The use of this information was proposed for the decoding of tailbiting codes in previous works \cite{ref6}. In that case, the soft-output Viterbi algorithm (SOVA) was employed to find a proper initial state for the trellis, achieving a fixed decoding time for all channel conditions with good performance in terms of both BER and BLER.

This work is organized as follows. In Section II a detailed explanation of the modifications introduced in the A* algorithm is given. A step-by-step description of the algorithm is also given at the end of the section. Section III presents the simulation results and the computational load of the proposed algorithm. Finally, in Section IV the main conclusions are summarized.

\section{Description of the Proposed Algorithm}

Let $C$ be a $(n, 1)$ binary tailbiting convolutional code generated with an encoder of memory $K$ and whose codewords are of length $nL$ bits. The trellis of this code $T$ spans $N = 2^K$ states at each time instant $i$, with $i$ ranging from $0$ to $L$. This trellis is formed by $N$ subtrellises, each of them starting and ending at one of the $N$ possible states. The trellis can also be seen as a subset of a more general trellis $T_s$ without constraints concerning the initial and final states of each path in it. In this sense, each path in the trellis $T$ is also a path in $T_s$, but not in reverse. Each path in the trellis is formed by the ordered sequence of states $\mathbf{s}_a = (s_a^0,s_b^1,\dots,s_a^L)$. Similar to \cite{ref4}, the metric $m$ used for the transition (branch) between the state $s_j$ at instant $i-1$ and state $s_k$ at instant $i$ is defined as:
\begin{equation}
m(s_{j}^{i-1},s_{k}^{i}) = \sum_{l=0}^{n-1}{(x_l\oplus y_{l+(i-1)n})\left|L_{l+(i-1)n}\right|}
\end{equation}   
where $n$ is the number of output coded bits per information symbol, $x_l$ is the output bit $l$ corresponding to that state transition and $y_{l+(i-1)n}$ and $L_{l+(i-1)n}$ are the hard decision bit and the log-likelihood ratio (LLR) of the received sequence at time instant $l+(i-1)n$ respectively. The LLR is defined as:
\begin{equation}
L_{l+(i-1)n} = \frac{P(y_{l+(i-1)n}=1|r_{l+(i-1)n})}{P(y_{l+(i-1)n}=0|r_{l+(i-1)n})}
\end{equation} 
where $r_{l+(i-1)n}$ is the soft $l+(i-1)n$ received bit. The first phase of the algorithm consists of a typical Viterbi decoding over the trellis $T_s$ with the initial state metrics set to zero. In each update of the algorithm the accumulated metric of the survivor path at state $s_j$ and time instant $i$, $M(s_j^i)$, is stored:
\begin{equation}
M(s_j^i) = \min_p(\Gamma(s_{p}^{i-1},s_{j}^{i}))
\end{equation}
where $\Gamma(s_{p}^{i-1},s_{j}^{i})$ is the accumulated metric of the path which merges in state $s_j$ at time $i$ through state $s_p$. We also store the term $\Delta(s_j^i)$, defined as:
\begin{equation}
\Delta(s_j^i) = \mathop{\min_q}\nolimits^{*}(\Gamma(s_{q}^{i-1},s_{j}^{i}))-\min_p(\Gamma(s_{p}^{i-1},s_{j}^i))
\label{eq:delta}
\end{equation}
where $\mathop{\min}\nolimits^{*}(\Gamma(s_{q}^{i-1},s_{j}^{i}))$ corresponds to the accumulated metric of the path with the second minimum metric which ends
at state $s_j$ at instant $i$. When the trellis calculation is finished, the survivor paths at each one of the $N$ \textit{final states} of the trellis are sorted and tracebacked. In the following, the term \textit{final state} will always refer to a state which is at the end of the trellis, while the term \textit{last state} will refer to the last state of a path which has not yet reached the end of the trellis.

If the survivor with the minimum metric is also tailbiting (its initial and final states are the same), the algorithm stops and this path is selected as the ML tailbiting path. If not, the survivor with the minimum metric which is also a tailbiting path is searched and its metric is stored in the variable $\rho$.
The survivor paths with accumulated metrics higher than $\rho$ are discarded. If none of the survivors is tailbiting, $\rho$ is set to $\infty$.

The non-discarded survivors are stored and their \textit{final states} are set as the initial nodes for the A* algorithm, which only operates on the trellis $T$. The evaluation function $f$ used in the A* algorithm is:
\begin{equation}
    f(\mathbf{s}_{a, j}^i) = g(\mathbf{s}_{a, j}^i) + h(\mathbf{s}_{a, j}^i)
\end{equation}
with $\mathbf{s}_{a, j}^i$ a path of length $i$ which ends at state $s_j$ in the
tailbiting subtrellis $a$. The function $g$ stores the accumulated metric from the initial state $s_a$ to state $s_j^i$:
\begin{equation}
    g(\mathbf{s}_{a, j}^i) = g(\mathbf{s}_{a, j}^{i-1}) + m(s_p^{i-1},s_j^i)
\end{equation}
where the value of $g$ for $i=0$ is equal to $0$. The algorithm mantains the paths $\mathbf{s}_{a, j}^i$ in a queue ordered by ascending values of their $f$ function.

The heuristic function $h$ represents an estimation of the distance from the state $s_j^i$ to the \textit{final state} of the subtrellis $s_a^L$. This function is similar to that proposed in \cite{ref4} and is defined
as $h(\mathbf{s}_{a, j}^i) = \max(0,M(s_a^L)-M(s_j^i))$ for $i$ greater than $0$.
For $i=0$, the value of $h(\mathbf{s}_{a, a}^0)$ is set to $M(s_a^L)$ if the zero-length path has never been on top of the queue or to $M'(s_a^L)$ if the zero-length path has been on top of the queue before. $M'(s_a^L)$ is defined as:
\begin{equation}
    M'(s_a^L) = M(s_a^L) + \min_{s_k^i} (\Delta(s_m^1), \Delta(s_n^2), \ldots \Delta(s_a^L))
    \label{eq}
\end{equation}

Second term in \eqref{eq} refers to the minimum $\Delta(s_k^i)$ corresponding
to the sequence of states which form the survivor path ending at state $s_a^L$ in the first phase of the algorithm. Thus, $ M'(s_a^L)$ corresponds to the accumulated metric of the second best path ending at the same \textit{final state} of that survivor. Since this definition of $h$ does not overestimate the actual distance from $s_j^i$ to $s_a^L$, the path found with the A* algorithm will have the mimimum metric, hence being optimal (ML).

In each iteration of the A* algorithm, the initial and \textit{last} states of the top-queued path are searched in a close table. If they have been previously recorded in it, the path is discarded. If not, these states are stored in the table and the successor paths of the top-queued path are obtained. The successor paths are formed by the concatenation of the top-queued path and its successor states (i.e., the states in the trellis that can be reached from the \textit{last state} of the top-queued path with a branch of length one). The values of the $f$ functions of the successor paths are then calculated. Finally, the top-queued
path is removed and its successor paths are inserted in the ordered queue.

It must be noted that if the branch between the \textit{last state} and a successor state of the top-queued path is the same as a surviving branch in the first phase of the algorithm, the calculation of the $f$ function can be avoided since it remains unchanged. Likewise, if the \textit{last state} of the top-queued path corresponds to a state of the stored survivor path computed in the first phase of the algorithm which ends at the initial state of the top-queued path, the algorithm stops. In this case, the path formed by the concatenation of the top-queued path and the rest of the survivor path from the state they join to its end is decoded. This proposed early stopping rule can be applied since the $f$ function of this resulting path will be the same as the $f$ function of the top-queued path.

The A* algorithm continues until the top-queued path reaches the \textit{final state} of its subtrellis, the queue is empty or the previous condition is fulfilled. The following steps summarize the proposed algorithm:

\begin{enumerate}
    \item \textit{Apply the Viterbi algorithm to the trellis $T_s$ with the initial state metrics set to zero. Record in each update the terms $M(s_j^i)$ and $\Delta(s_j^i)$ for all the states. Record also the survivor paths at the end of the algorithm.}
    \item \textit{If the survivor path with the minimum metric is also tailbiting (its initial and final state are the same), the algorithm stops and this path is decoded.}
    \item \textit{Set $\rho$ as the minimum metric of the survivor path which is also tailbiting. If none of the survivors is tailbiting, $\rho$ is set to $\infty$.}
    \item \textit{Discard all the survivor paths with metrics $M(s_j^L)$ higher than $\rho$.}
    \item \textit{Load in the queue the zero-length paths corresponding to the final states of the paths not discarded in the previous step. Sort them in ascending order of their $f$ function values, which correspond to $M(s_j^L)$.}
    \item \textit{If the queue is empty, the algorithm stops and the survivor found in the first phase of the algorithm with accumulated metric $\rho$ is decoded.}
    \item \textit{If the top-queued path reaches the end of its subtrellis, the top-queued path is decoded.}
    \item \textit{If the top-queued path merges into the survivor path found in the first phase of the algorithm which ended in the final state of its subtrellis, the algorithm stops and the path formed by the concatenation of the top-queued path and the rest of the survivor path from the state they join is decoded.}
    \item \textit{If the top-queued path has zero length and its $f$ function value corresponds to $M(s_j^L)$, change this value to $M'(s_j^L)$. If $M'(s_j^L)$ is greater than $\rho$, the path is discarded. If not, rearrange the queue in ascending order of the $f$ function and go to step 6.}
    \item \textit{If the initial and final states of the top-queued path are stored in the close table, discard the path and go to step 6. If not, store them in the close table.}
    \item \textit{Find the successors of the top-queued path and compute their $f$ functions. Delete the top-queued path and rearrange the queue in ascending order of the $f$ function. If some path has an $f$ function higher than $\rho$, delete it. Go to step 6.}
\end{enumerate}

\section{Simulation Results}
In this section, we show the performance of the proposed decoding algorithm, which we will call EA* henceforth, over the additive white Gaussian noise (AWGN) channel. The considered code is the $(96,48)$ tailbiting code with generator
polynomials $(171,133)$ in octal. This coding scheme is used in the 802.16e standard. The encoded data are mapped to QPSK symbols and transmitted. The results were obtained ensuring at least 100 reported errors for every simulation result.

Table \ref{table} compares the complexity of the proposed algorithm with that of IA* in terms of $f$ function calculations and total number of operations. In this sense, each iteration of the A* algorithm may require:
\begin{enumerate}
    \item 1 comparison to check if the top-queued path has reached the end of its subtrellis. 
    \item 1 search in the close table: $O(\log_2(C))$ comparisons, with $C$ the size of the table.
    \item 1 comparison per successor to check if it is required the calculation of the $f$ function.
    \item 2 additions (calculation of $f$ and $g$) and 1 subtraction (calculation of $h$) per successor to compute the $f$ function.
    \item 1 search in the ordered queue per successor: $O(\log_2(Q))$ comparisons, with $Q$ the size of the queue.
\end{enumerate}

\begin{table}[ht]
\caption{Mean $f$ function calculations and total number of operations of IA* and EA* in the second phase of the algorithm}
\renewcommand{\arraystretch}{1.3}
\centering
\begin{tabular}{|c||c|c|c|c|}
\hline
$E_b/N_0$ & 2dB & 3dB & 4dB & 5dB \\
\hline
\multicolumn{5}{|c|}{Mean $f$ function calculations} \\
\hline
IA* & 295 & 76 & 24 & 11 \\ \hline
EA* & 177 & 35 & 9 & 3 \\ \hline
\multicolumn{5}{|c|}{Mean total number of operations} \\ \hline
IA* & 6623 & 1634 & 556 & 332 \\ \hline
EA* & 4201 & 814 & 226 & 114 \\ \hline
\end{tabular}
\label{table}
\end{table}

Apart from these operations, EA* requires per iteration of the A* algorithm an additional comparison to check if the top-queued path has merged with the ML path and $L$ comparisons plus 1 addition if the top-queued path has never been before on top of queue and its length is zero. Concerning the first phase of the algorithm, the computational load of the required subtraction in \eqref{eq:delta} when the Viterbi algorithm is being performed can be neglected since it substitutes the comparison required in each Viterbi update (the comparison can be easily done considering the sign of the subtraction). As can be seen in Table \ref{table}, the decrease in the number of required operations when EA* is used ranges from 36\% for 2dB to 64\% for 5dB.

With regard to the memory requirements, the EA* stores the non discarded survivor paths found in the first phase of the algorithm and their associated metric differences. Nevertheless, since the number of entries in the ordered queue and the close table is lower than that of the IA*, the total required memory is similar in both algorithms.

\section{Conclusion}
A new ML decoding algorithm for tailbiting codes based on the A* algorithm is proposed. The algorithm first performs a Viterbi decoding to obtain information of the trellis. Later, this information is used in the A* algorithm to compute the
heuristic function. Simulation results show that the proposed algorithm achieves a marked decrease in the complexity of the overall decoding process.

\bibliographystyle{IEEEtran}

\end{document}